\newif\ifarxiv
\arxivtrue % or
\ifarxiv
	\documentclass[12pt]{article}
	\usepackage[margin=1in]{geometry}
	\usepackage[slantedGreek]{mathpazo}
	\usepackage[pdftex]{graphicx}
	\usepackage{amsfonts}
	\usepackage{amssymb}
	\usepackage{amsthm}
	\usepackage{graphicx,float}
	\usepackage{amsmath}%
	\usepackage{hyperref}
	\usepackage{natbib}
\else
\fi
\usepackage[utf8]{inputenc}
\usepackage{color,soul,marginnote}
\usepackage{array}
\usepackage{breakcites}
\usepackage{breakcites}

\DeclareMathOperator*{\minf}{minimize \quad}

\newcommand{\mininlineeq}[4]{\begin{equation}\label{#4}\mbox{minimize}_{#1}\quad#2\qquad\mbox{subject to }#3\end{equation}}

\ifarxiv
\title{Bayesian Regularization: From Tikhonov to Horseshoe}
\author{
	Nicholas G. Polson\thanks{Nicholas G. Polson is a Professor of Econometrics and Statistics at The University of Chicago Booth School of Business} ~and Vadim Sokolov\thanks{Vadim Sokolov is Assistant Professor in Operations Research at George Mason University. email:vsokolov@gmu.edu}}
\date{First Draft: November, 2018\\
	This Draft: February, 2019
	}
\else
\fi

\graphicspath{{./fig/}}

\begin{document}
\ifarxiv
\maketitle
\begin{abstract}
\noindent
Bayesian regularization is a central tool in modern-day statistical and machine learning methods. Many applications involve high-dimensional sparse signal recovery problems. The goal of our paper is to provide a review of the literature on  penalty-based regularization approaches, from Tikhonov (Ridge, Lasso) to horseshoe regularization.
\end{abstract}
\else
\fi

\section{Introduction}
Regularization is a machine learning technique that allows for an optimal trade-off between model complexity (bias) and out-of-sample performance (variance). To fix ideas, consider regularization in the context of a linear model, where an output $y$ is generated by
\begin{equation}\label{eq:lm}
y = x^T\beta + \epsilon,~~~\epsilon\sim p(\epsilon).
\end{equation}
Assuming normally distributed errors, $p(\epsilon) = N(0,\sigma^2_{\epsilon})$, the corresponding regularized maximum likelihood optimization problem is  finding the solution to
\mininlineeq{\beta}{||y- X\beta||^2_2 }{\sum_{i=1}^{p}\phi(\beta_i) \le s.}{eq:regopt}
Here, $y$ is the vector of observed outputs, $X$ is a design matrix, and $\beta$ are the model parameters. Each $\beta_i$ has a regularization penalty $\phi(\beta_i)$ and $s$ is a hyper-parameter that controls the bias-variance trade-off. 

Regularization can be viewed as constraint on the model space. The techniques were originally applied to solve ill-posed problems where a slight change in the initial data could significantly alter the solution.  Regularization techniques were then proposed for parameter reconstruction in a physical system modeled by a linear operator implied by a set of observations.  It had long been believed that ill-conditioned problems offered little practical value, until Tikhonov published his seminal paper~\citep{tikhonov1943stability} on regularization.  \cite{tihonov1963solution} proposed methods for solving regularized problems of the form
\[
\minf_\beta ||y- X\beta||^p_p   + \lambda||(\beta - \beta^{(0)})||^q_q.
\]
Here $\lambda$ is the weight on the regularization penalty and the $\ell_q$-norm is defined by $||\beta||_q^q = \sum_i \beta_i^q$. This optimization problem is a Lagrangian form of the constrained problem given in Equation~(\ref{eq:regopt}) with $\phi(\beta_i) = (\beta_i - \beta_i^{(0)})^q$.

The subsequent developments were proposed in~\cite{ivanov1962linear} and numerical algorithms were then developed by \cite{bakushinskii1967general}. All of these methods required developing approximations by well-posed problems, parameterized by the regularization parameter. Most of the early work in Soviet literature focused on proving convergence of the solutions of well-posed problems to the ill-posed problems. Numerical schemes were proposed much later. For a detailed overview of earlier convergence and numerical results, see~\cite{tikhonov1977methods} and \cite{ivanov2013theory}.

In the context of linear models in statistics~\cite{hoerl1970ridge} derived statistical properties of regularized estimators in case when penalty has $\ell_2$ norm and $p=q=2$. This estimator was called the Ridge regression. 

Later, sparsity became a primary driving force behind new regularization methods~\cite{candes2008introduction}.  When the penalty term has $\ell_1$ norm ($p=2,~q=1$), the solution to regularized problem is sparse, e.g. has many zeros~\citep{alliney1992digital,donoho1992superresolution,donoho1995adapting,aster2018parameter}. Use of $\ell_0$~\citep{polson2017bayesian} pseudo-norm, which counts the number of non-zero entries in a vector, leads to a NP hard optimization problem. $\ell_1$ penalty can be viewed as a convex approximation of $\ell_0$ penalty which still has the required property of recovering sparse vectors of parameters. An algorithm for estimating $\ell_1$ regularized linear statistical model was proposed by~\cite{alliney1994algorithm}. \cite{williams1995bayesian} used Bayesian approach that assigns Laplace prior for parameters of non-linear neural network models.~\cite{tibshirani1996regression} derived statistical properties of $\ell_1$ regularization based estimators for linear models and coined the term lasso for this problem. For brief historical accounts on the use of the $\ell_1$ penalty in statistics and signal processing, see \cite{tibshirani1996regression,miller2002subset}, and the total variational denoising literature~\cite{claerbout1973robust,taylor1979deconvolution}.

\section{Bayesian Regularization: From Tikhonov to Horseshoe}
Mathematically, one can to think of defining a regularized solution by constraining the topology of a search space to a ball. From a  Bayesian perspective instead assigns a prior distribution to each of the model's parameters. From a historical perspective, James-Stein (a.k.a $L^2$-regularization)~\cite{stein1964inadmissibility} provided a global shrinkage rule for improving statistical estimation. There are no local parameters to learn about sparsity, which led to horseshoe regularization. 

\subsection{Bayes Risk}
A simple sparsity example illustrates the issue with $L^2$-regularization and the James-Stein estimator. Consider  the sparse $r$-spike  problem and focus solely on rules with the same shrinkage weight (albeit benefiting from pooling of information). Let the true parameter value be $ \theta_p = \left ( \sqrt{d/p} , \ldots , \sqrt{d/p} , 0 , \ldots , 0 \right ) $.
James-Stein is equivalent to the model
$$
y_i = \theta_i + \epsilon_i \; {\rm and} \; \theta_i \sim \mathcal{N} \left ( 0 , \tau^2 \right ) 
$$
This dominates the plain MLE but loses admissibility because a ``plug-in'' estimate of global shrinkage $\hat{\tau}$ is used. Original ``closed-form'' analysis is particularly relevant here~\citep{tiao1965bayesian}. They point out that the mode of $p(\tau^2|y)$ is zero exactly when the shrinkage weight turns
negative (their condition 6.6). From a risk perspective $ E \Vert \hat{\theta}^{JS} - \theta \Vert \leq p , \forall \theta $ showing the inadmissibility of the  MLE. At origin the risk is $2$, but
$$
\frac{p \Vert \theta \Vert^2}{p + \Vert \theta \Vert^2} \leq R \left ( \hat{\theta}^{JS} , \theta_p \right ) \leq
2 + \frac{p \Vert \theta \Vert^2}{ d + \Vert \theta \Vert^2}.
$$
This implies that $ R \left ( \hat{\theta}^{JS} , \theta_p \right ) \geq (p/2) $.
Hence, simple thresholding rule beats James-Stein this with a risk given by $ \sqrt{\log p } $. This simple example, shows that the choice of penalty should not be taken for granted as different estimators will have different risk profiles.

\subsection{Bayesain Regularization Duality}
From a Bayesian perspective regularization is performed by defining a prior distribution over the model parameters. A Bayesian linear regression model is defined as
\begin{equation}
y =  x^T\beta+ \epsilon,~~\epsilon\sim N(0,\sigma^2_{\epsilon}),~~~\beta\sim p(\beta\mid \tau),
\end{equation}
the log of the posterior distribution is then given by 
\[- \log p(\beta\mid X,y) =  (1/2) \sigma_{\epsilon}^2 \sum_{i} (y_i - x_i^T\beta)^2 + \log p(\beta\mid \tau).\]

A regularized maximum a posteriori probability (MAP) estimator can be found by minimizing the negative log-posterior
\begin{equation}
\label{eqn:reg}
\begin{aligned}
& \hat \beta_{\mathrm{MAP}} = \underset{\beta \in R^p}{\text{argmin}}
& & ||y-X\beta||_2^2+ \phi_{\tau}(\beta),
\end{aligned}
\end{equation}
where $\phi_{\tau}(\beta)\propto \log  p(\beta \mid \tau) $. The penalty term is interpreted as  the log of the prior distribution,  and is parametrized by the hyper-parameters $\tau$.  The resulting maximum a posteriori probability (MAP)  is equivalent to the classical approach of constraining a search space by adding a penalty.

Table~\ref{tab:prior-penalty} provides penalty functions and their corresponding prior distributions, including lasso, ridge, Cauchy and horseshoe. 
\begin{table}[H]
	\begin{tabular}{c|cccc}
		& Ridge & Lasso & Cuachy & Horseshoe \\ \hline
		Prior $p(\beta_i\mid \tau)$ & $\frac{1}{\sqrt{2 \pi } \tau }\exp\left(-\frac{\beta_i^2}{2 \tau ^2}\right)$ & $\frac{1}{2 \tau }\exp\left(-\frac{|\beta|}{\tau}\right)$ & $\frac{\tau }{\pi  \tau ^2+\pi  \beta_i^2}$ & $\le \pi\sqrt{\dfrac{\pi}{2}}\log \left ( 1 + \frac{2 \tau^2}{\beta_i^2} \right )$ \\ 
		Penalty $\phi_{\tau}(\beta_i)$ & $\dfrac{1}{2\tau^2}\beta_i^2$ & $\frac{|\beta_i|}{\tau}$ & $\log\left( \tau ^2+ \beta_i^2\right)$ & $- \log \log \left ( 1 + \frac{2 \tau^2}{\beta_i^2} \right )$  \\ 
	\end{tabular} 
	\label{tab:prior-penalty}
	\caption{Prior distributions and corresponding penalty functions (negative log-probability)}
\end{table}

\begin{figure}[H]
	\begin{tabular}{ccccc}
		&$\ell_2$ (Ridge) & $\ell_1$ (Lasso) & Cauchy & Horseshoe \\
		\raisebox{1\height}{\rotatebox{90}{unit ball}}&\includegraphics[width=0.23\linewidth]{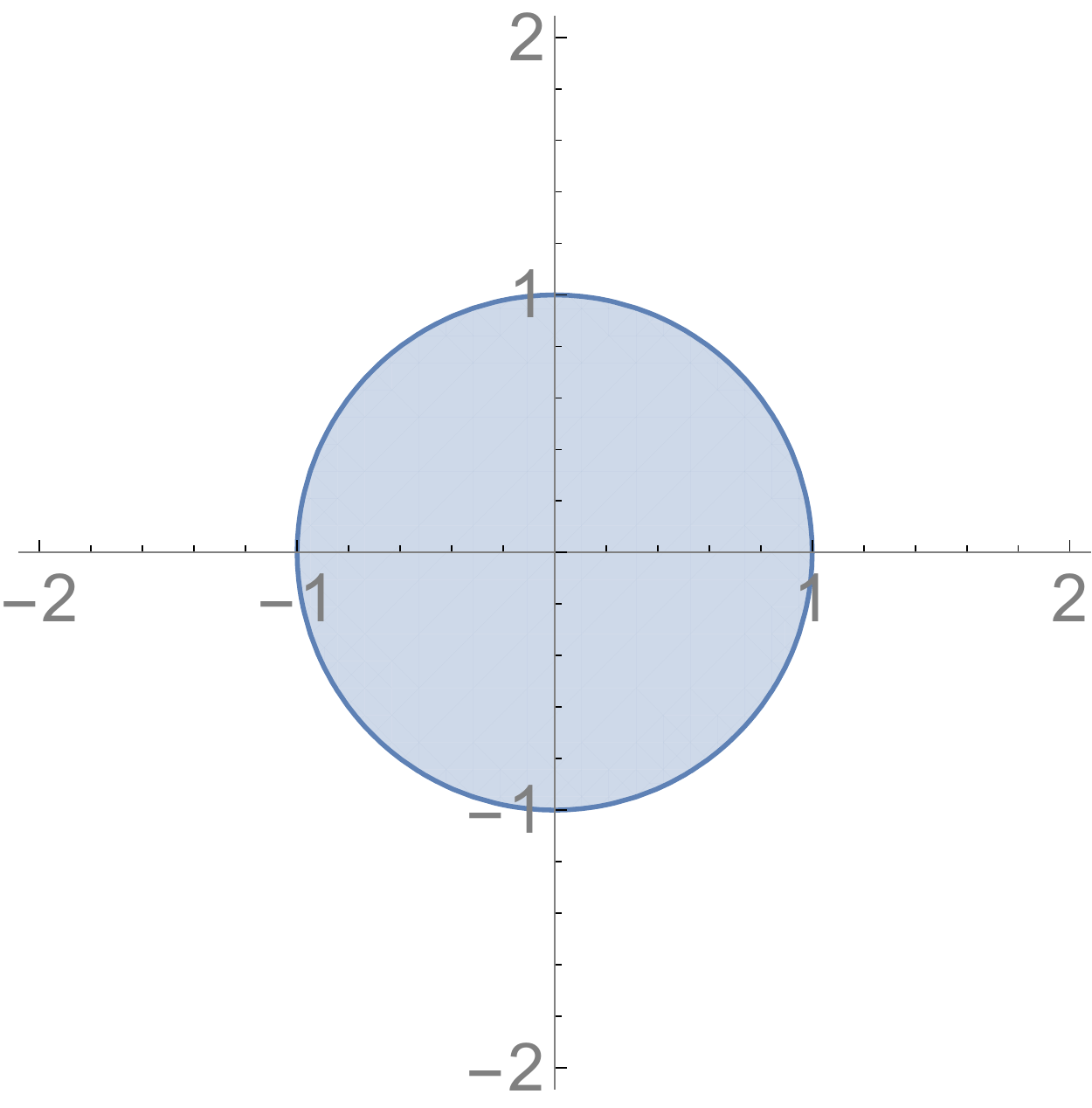} & \includegraphics[width=0.23\linewidth]{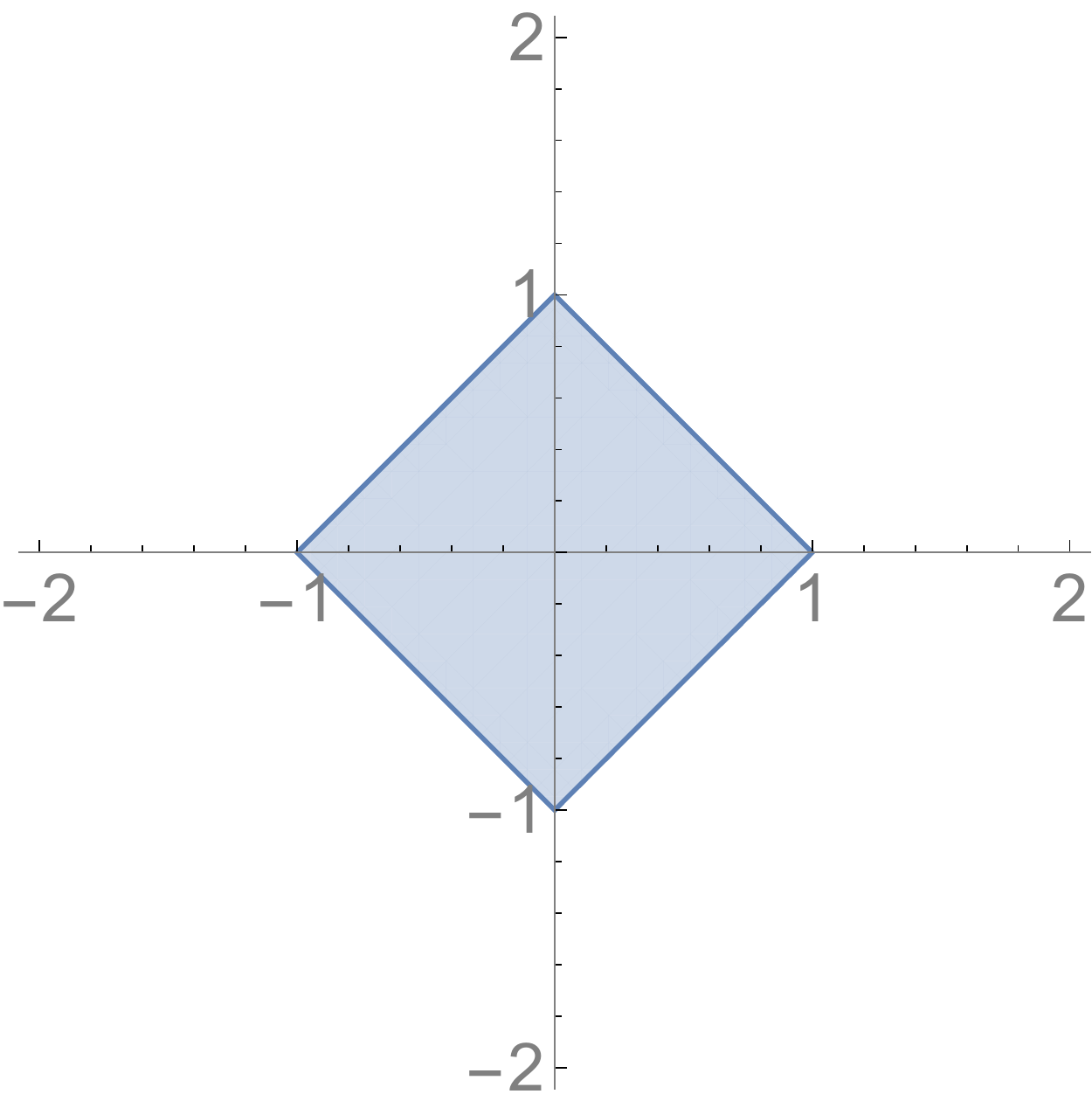} &
		\includegraphics[width=0.23\linewidth]{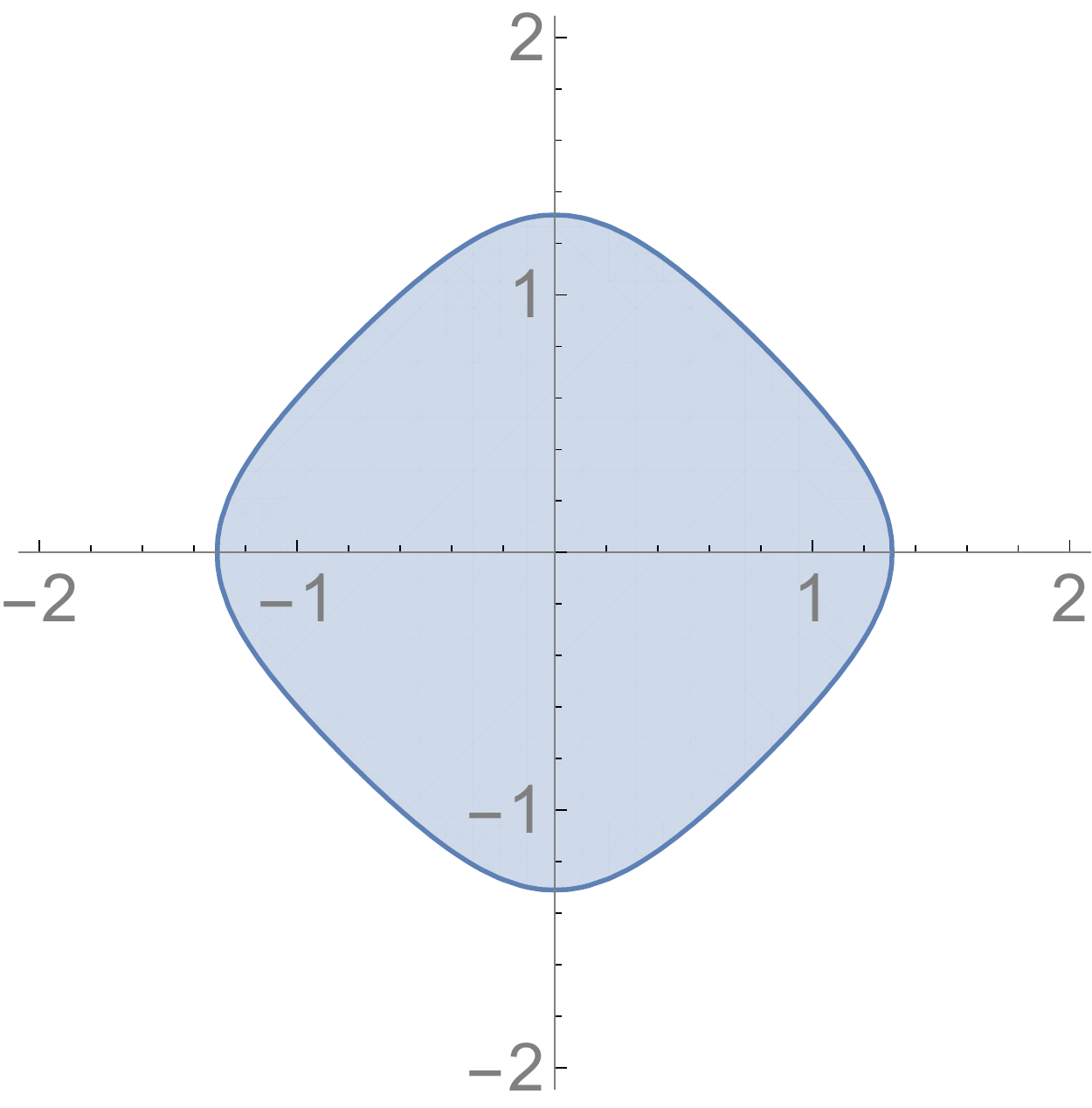} & \includegraphics[width=0.23\linewidth]{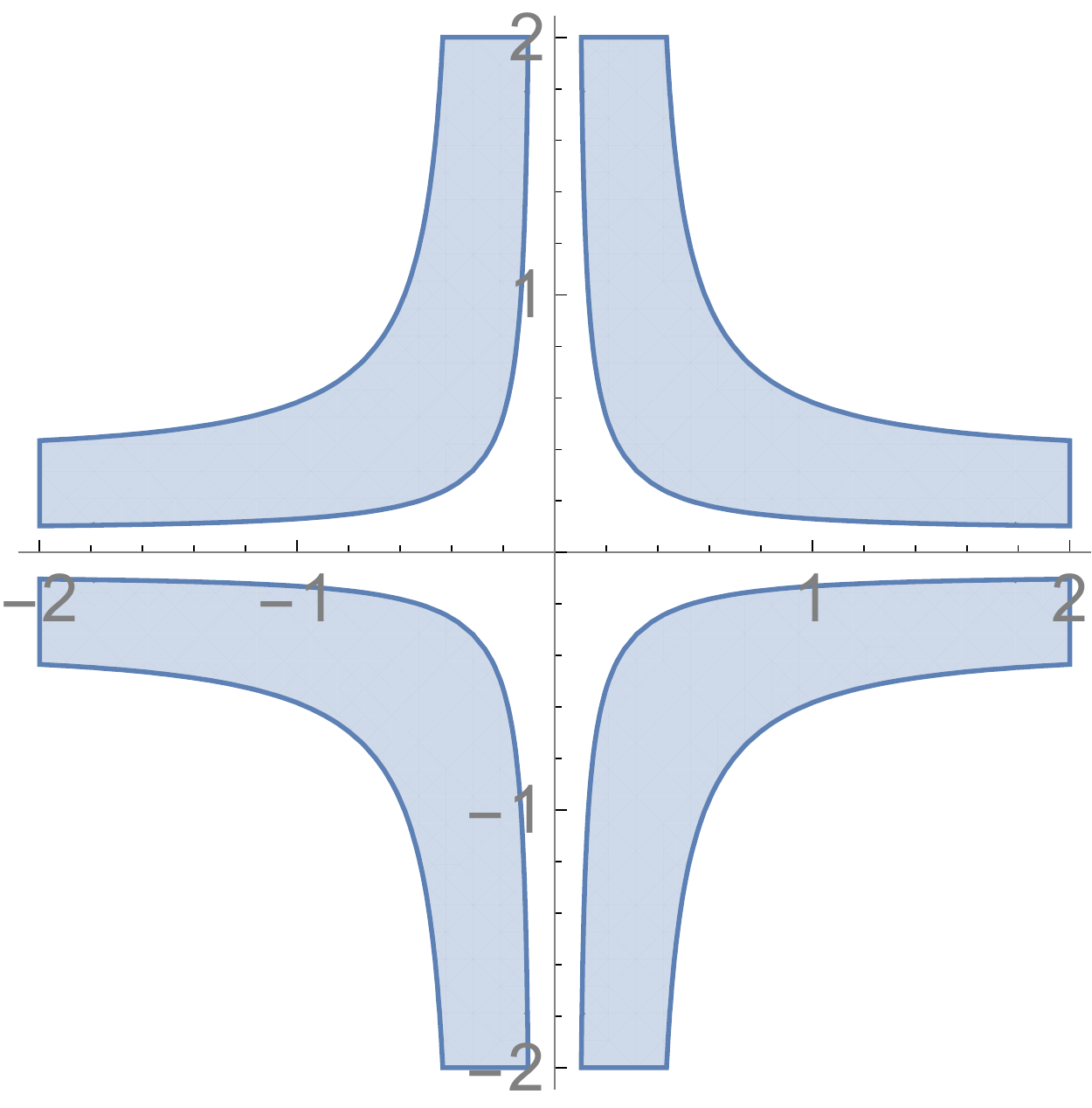}\\
		&Normal & Laplace & Cauchy & Horseshoe \\
		\raisebox{1\height}{\rotatebox{90}{prior}} & \includegraphics[width=0.23\linewidth]{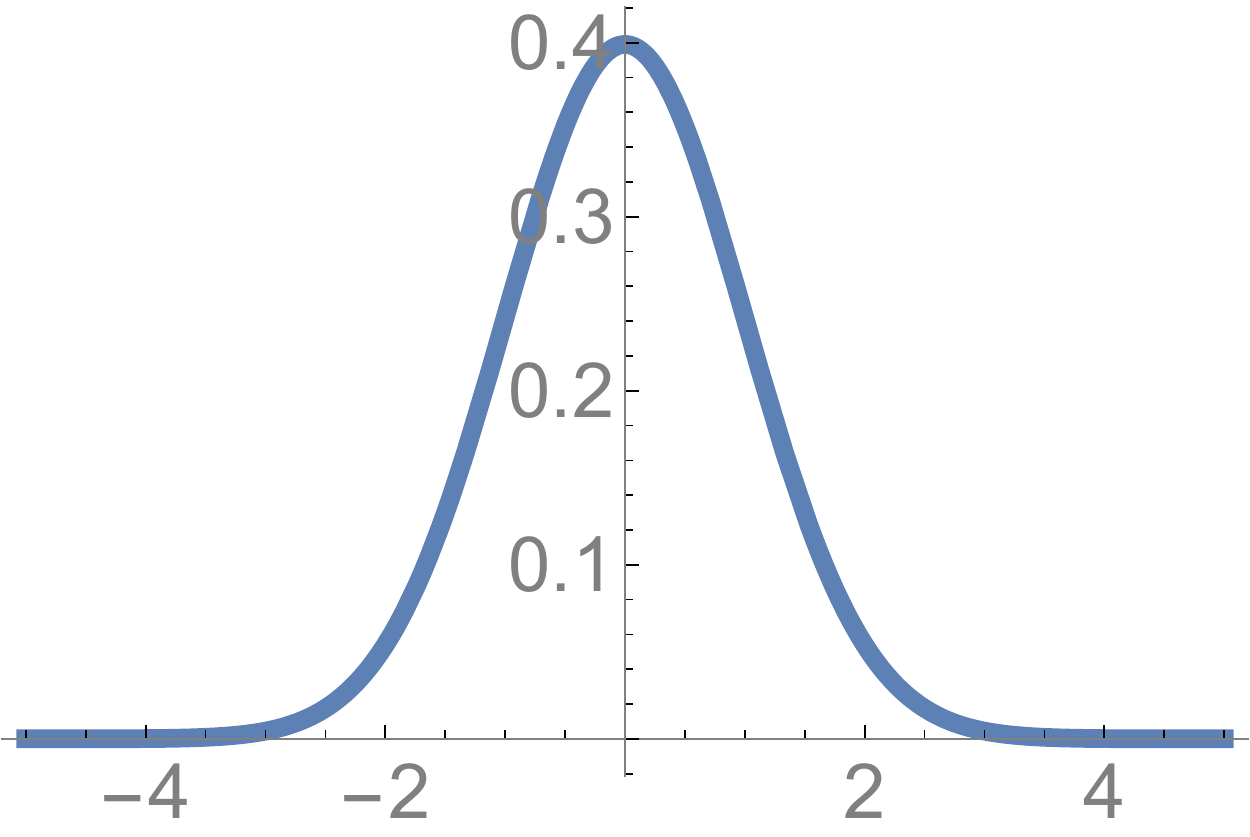} & \includegraphics[width=0.23\linewidth]{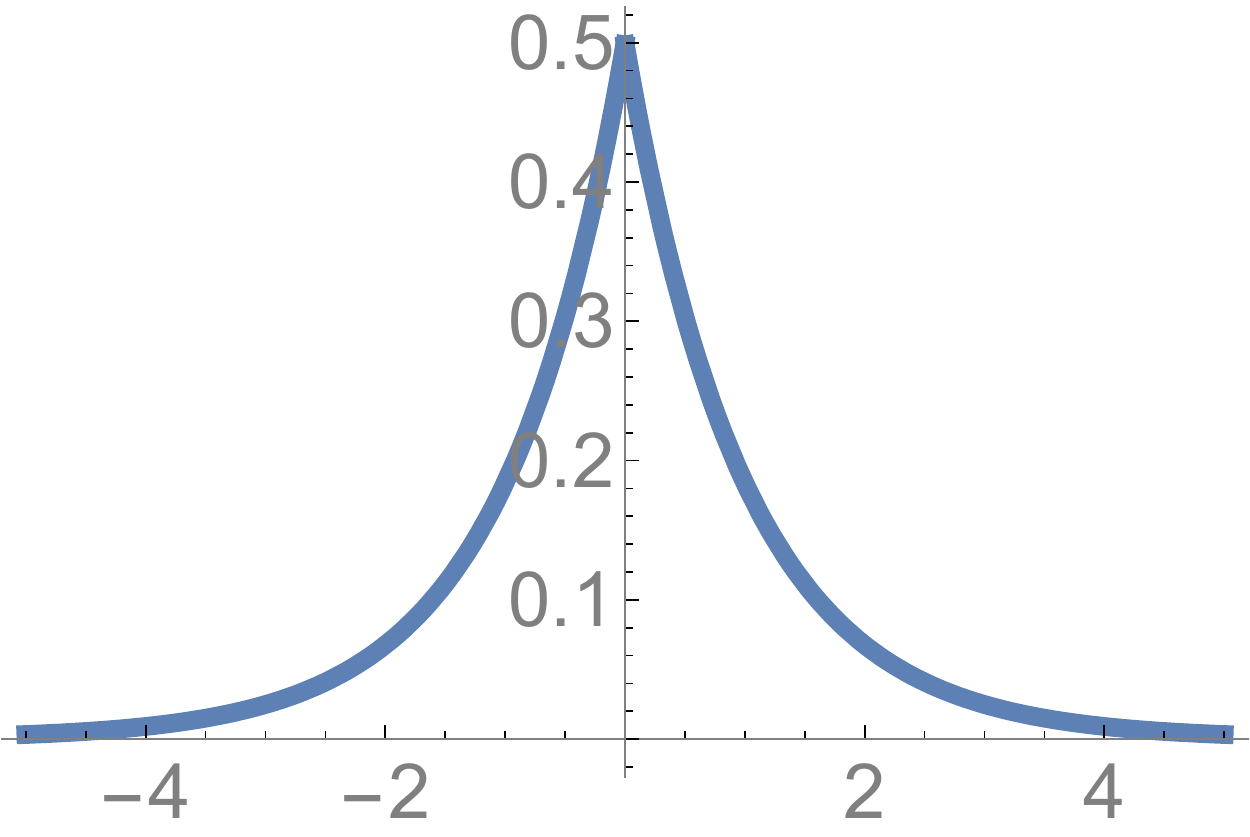} &
		\includegraphics[width=0.23\linewidth]{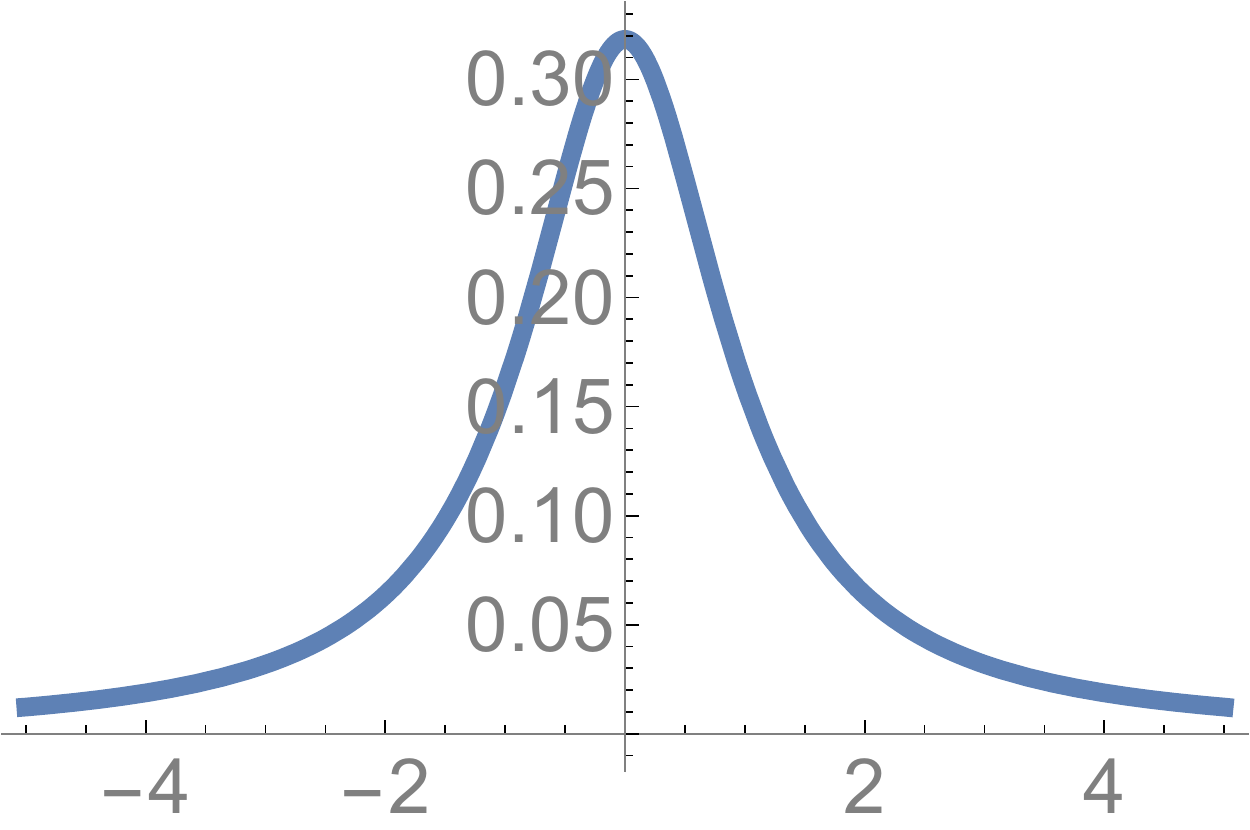} & \includegraphics[width=0.23\linewidth]{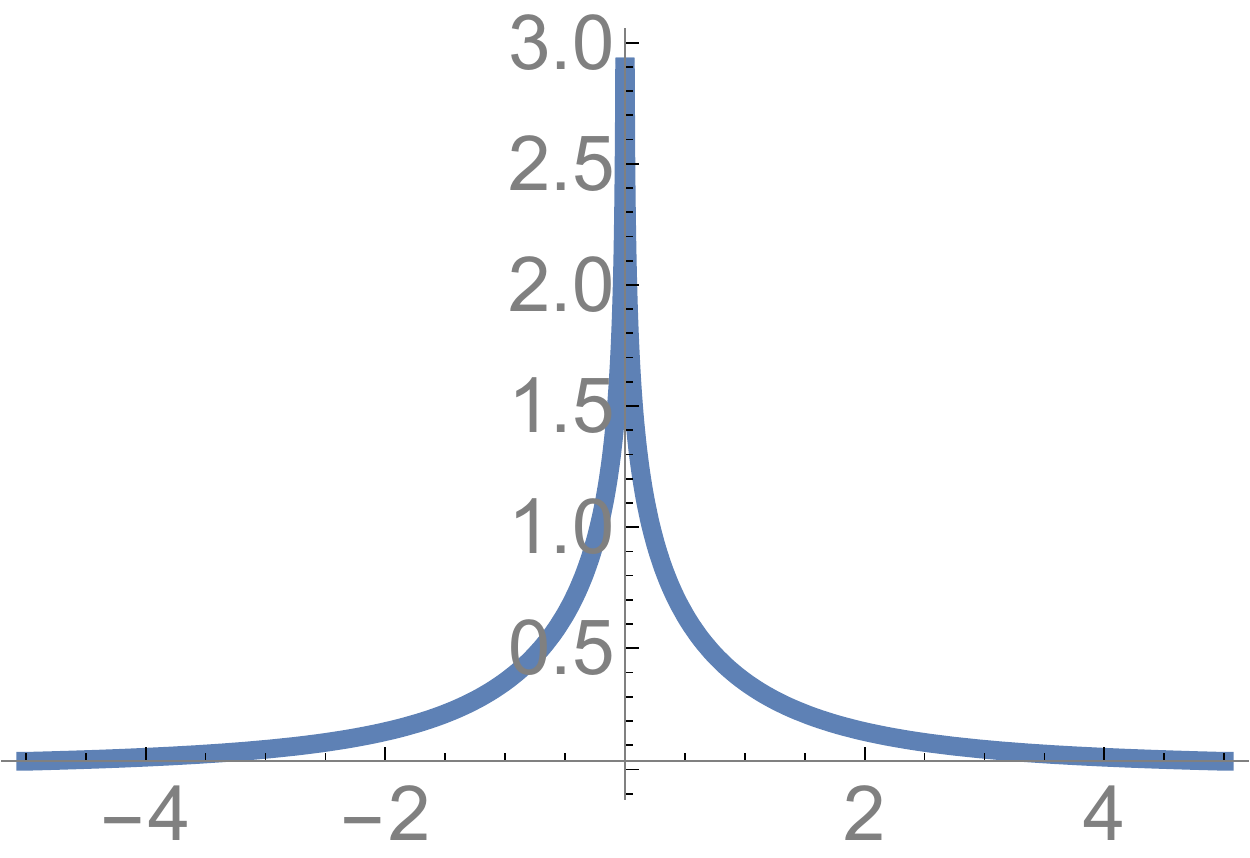}
	\end{tabular}
	\caption{Comparison of geometry of a unit ball induced by Normal, Laplace, Cauchy and Horseshoe priors.}
	\label{fig:pealty}
\end{figure}

Figure~\ref{fig:pealty} compares the geometry of a unit ball which is used as a constraint in traditional approach and the corresponding prior distribution as used in Bayesian approach, we show ridge, lasso, Cauchy, and horseshoe penalties. 

%In Lasso and Ridge the negative log-likleihood defines the loss function  $l\left(\beta\right) = -\log p(y\mid \theta,x)$ and log-prior defines a penalty function, denoted by $\phi_{\lambda}(\beta) = \log p(\theta)$, where $\lambda$ is a global regularization parameter.  
%
%From Bayesian perspective regularization is nothing but incorporation of prior information into the model. 

A typical approach in Bayesian analysis is to define normal scale mixture priors which are constructed as a hierarchical model of the form
\begin{equation}\label{eq:scalemix}
\beta_i \mid \lambda_i,\tau_i \sim N(0,\tau_i^2\lambda_i^2),~~~p(\lambda^2,\sigma^2,\tau) = p(\lambda^2)p(\sigma^2)p(\tau)
\end{equation}

While classical approach requires solving an optimization problem, the Bayesian approach requires calculating integrals. While in conjugate models, e.g. when both likelihood and priors are normal (Ridge), we can calculate those integrals analytically, it is not possible in general case. 
An efficient numerical techniques for calculating samples from posterior distributions are required. \cite{george1993variable} proposed a Gibbs sample for finding posterior of the following problem 
\[
\beta_i\mid \gamma_i \sim (1-\gamma_i)N(0,\tau^2_i) + \gamma_iN(0,c_i^2\tau_i^2),~~~ p(\gamma_i=1) = p_i,
\]
where $\tau_i$ is chosen to be small, so that for $\gamma_i = 0$, the estimated $\beta_i$ is close to zero and and $c_i$ is large so that when $\gamma_i = 1$ the estimated $\beta_i$ does not get shrunk. Then variable selection is performed by calculating the posterior distribution over $\gamma$.
\[
p(\gamma\mid X,y ) \propto p(y\mid X,\gamma)p(\gamma).
\]

\cite{carlin1991inference} proposed Gibbs sampling MCMC for the class of scale mixtures of Normals, taking the form
\[
\epsilon_j \mid \sigma,\lambda_j \sim N(0,\lambda_j\sigma^2),~~~\lambda_j \sim p(\lambda_j)
\]
We now turn to lasso and horseshoe as special cases. 
%\cite{carlin1992monte} developed Monte Carlo sampling algorithms for time series models.

\subsection{Lasso}
From a Bayesian perspective, lasso~\citep{tibshirani1996regression} is equivalent to  specifying double exponential (Laplace) prior distribution~\cite{carlin1991inference} for each parameter  $\beta_i$ with $\sigma^2$ fixed
\[
p(\beta_i \mid b) = (b/2)\exp(-|\beta_i|/b).
\]
Bayes rule then calculates  the posterior as a product of Normal likelihood and the Laplace prior to yield
\[
\log p(\beta \mid X,y, b) \propto ||y-X\beta||_2^2 + \dfrac{2\sigma^2}{b}||\beta||_1.
\]
For $b>0$, the posterior mode is equivalent to the LASSO estimate with $\lambda = 2\sigma^2/b$. Large variance $b$ of the prior is equivalent to the small penalty weight $\lambda$ in the Lasso objective function.

The Laplace prior used in Lasso can be  represented as scale mixture of Normal distribution~\citep{andrews1974scale,carlin1991inference}
\begin{align*}
\beta_i \mid \sigma^2,\tau \sim &N(0,\tau^2\sigma^2)\\
\tau^2  \mid \alpha \sim &\exp (\alpha^2/2)\\
\sigma^2 \sim & \pi(\sigma^2).
\end{align*}
There is an equivalence with the lasso penalty obtained by integrating out $\tau$
\[
p(\beta_i\mid \sigma^2,\alpha) =  \int_{0}^{\infty} \dfrac{1}{\sqrt{2\pi \tau}}\exp\left(-\dfrac{\beta_i^2}{2\sigma^2\tau}\right)\dfrac{\alpha^2}{2}\exp\left(-\dfrac{\alpha^2\tau}{2}\right)d\tau = \dfrac{\alpha}{2\sigma}\exp(-\alpha/\sigma|\beta_i|).
\]
Thus it is a Laplace distribution with location 0 and scale $\alpha/\sigma$.

\cite{carlin1991inference,carlin1992monte,park2008bayesian} used representation of Laplace prior is a scale Normal mixture to develop a Gibbs sampler that iteratively samples from $\beta \mid a,y$ and $b\mid \beta,y$ to estimate joint distribution over $(\hat \beta, \hat b)$. Thus, we so not need to apply cross-validation to find optimal value of $b$, the Bayesian algorithm does it ``automatically''. 
Given data $D = (X,y)$, where $X$ is the  $n\times p$ matrix of standardized regressors and $y$ is the $n$-vector of outputs.  Implement a Gibbs sampler for this model when Laplace prior is used for model coefficients $\beta_i$. Use scale mixture normal representation.
\begin{align*}
\beta \mid  \sigma^2,\tau_1,\ldots,\tau_p \sim  & N(0,\sigma^2D_{\tau})\\
D_{\tau} = & \mathrm{diag}(\tau_1^2,\ldots,\tau_p^2)\\
\tau_i^2  \mid \lambda \sim &\exp (\lambda^2/2)\\
\sigma^2 \sim & 1/\sigma^2.
\end{align*}

Then the complete conditional required for Gibbs sampling are given by
\begin{align*}
\beta \mid D,D_{\tau} \sim & N(A^{-1}X^Ty,\sigma^2A^{-1}), ~~A = X^TX + D^{-1}_{\tau}\\
\sigma^2 \mid \beta,D,D_{\tau} \sim & \mathrm{InverseGamma}\left((n-1)/2+p/2,(y-X\beta)^T(y-X\beta)/2 + \beta^TD_{\tau}^{-1}\beta/2\right)\\
1/\tau_j^2 \mid \beta_j,\lambda \sim & \mathrm{InverseGaussian}\left(\sqrt{\dfrac{\lambda^2\sigma^2}{\beta_j^2}},\lambda^2\right)
\end{align*}
The formulas above assume that  $X$ is standardized, e.g. observations for each feature are scaled to be of mean 0 and standard deviation one, and $y$ is centered $y = y - \bar y$. 

You can use empirical priors and initialize the parameters as follows
\begin{align*}
\beta = & (X^TX + I)^{-1}X^Ty\\
r = & y - X\beta\\
\sigma^2 = & r^Tr/n\\
\tau^{-2} = &1/(\beta \odot \beta)\\
\lambda  = &  p  \sqrt{\sigma^2} / \sum|\beta|.
\end{align*}
Here $n$ is number of rows (observations) and $p$ is number of columns (inputs) in matrix $X$.

\subsection{Ridge}
When prior is Normal $\beta_i \sim N(0,\sigma_{\beta}^2)$, the posterior mode is equivalent to the ridge~\cite{hoerl1970ridge} estimate. The relation between variance of the prior  and the penalty weight in ridge regression is inverse proportional $\lambda\propto 1/\sigma_{\beta}^2$.

Thus, Lasso and Ridge regressions are both maximum a posteriori (MAP) estimates for Laplace and Normal priors.

Given design matrix $X$ and observed output values $y = (y_1,\ldots,y_n)$, and assuming $\epsilon \sim N(0,\sigma^2)$, the MLE  is given by the solution to the following optimization problem
\[
\operatornamewithlimits{minimize}_{\beta}\quad||X\beta^T - y||_2^2 
\]
and the solution is given by:
\[
\beta = \left(X^TX\right)^{-1}X^Ty.
\]
However, when matrix $X$ is close to being rank-deficient, the $X^TX$ will be ill-conditioned. This means that the problem of estimating $\beta$ will also be ill-conditioned. For a linear model, we can quantify the sensitivity to perturbation in $y$ by
\[
\dfrac{||\Delta \beta||}{||\beta||} \le \dfrac{\kappa(X^TX)}{\cos\theta}\dfrac{||\delta X^Ty||}{||X^Ty||},
\] 
here $\theta$ is the angle between $X^Ty$ and the range of $X^TX$ and $\kappa(X^TX)$ is the condition number which is the ratio of the largest to smallest eigenvalues  of $X^TX$. 

A trivial example is shown when $y$ is nearly orthogonal to $x$
\[
x = \left[\begin{array}{c}
1 \\ 
0
\end{array} \right], \ 
y^{(1)} = \left[\begin{array}{c}
\epsilon \\ 1
\end{array}\right]
\]
The solution to the problem is $\beta^{(1)} = \epsilon$; but the solution for
\[
x = \left[\begin{array}{c}
1 \\ 
0
\end{array} \right], \ 
y^{(2)} = \left[\begin{array}{c}
-\epsilon \\ 1
\end{array}\right]
\]
is $\hat\beta^{(2)} = -\epsilon$. Note that  $||y^{(1)}- y^{(2)}||/||y^{(1)}|| \approx 2\epsilon$ is small, but
$|\hat{\beta}^{(1)} - \hat{\beta}^{(2)}|/|\hat{\beta}^{(1)}| = 2$, is huge.

Another case of interest is when a least squares problem is ill-conditioned is when the observations are close to be linearly dependent. It happens, for example, when input variables are correlated. Consider an example
\[
X = \left(
\begin{array}{cc}
1 & 1 \\
1 & \epsilon +1 \\
\end{array}
\right),~~~y = \left(
\begin{array}{c}
2\\\delta+2
\end{array}
\right)
\]
The MLE estimate is given by
\[
\beta = \left\{2-\frac{\delta }{\epsilon },\frac{\delta }{\epsilon }\right\}
\]
For $\delta = 0$, we have $\hat \beta^{(1)}= (2,0)$ but for $\delta = \epsilon$, we have $\hat \beta^{(2)} = (1,1)$ with both $\epsilon$ and $\delta$ being arbitrarily small. We can analytically calculate the condition number
\[
\kappa(X^TX) = \frac{\epsilon ^2+(\epsilon +2) \sqrt{\epsilon ^2+4}+2 \epsilon +4}{\epsilon
	^2-(\epsilon +2) \sqrt{\epsilon ^2+4}+2 \epsilon +4}
\]
It goes to infinity as $\epsilon$ goes to zero. Since condition number is the ratio of eigenvalues
\[
\kappa(X^TX) = \dfrac{\lambda_{\mathrm{\max}}(X^TX)}{\lambda_{\mathrm{\min}}(X^TX)}
\]
and in our case $\lambda_{\mathrm{\min}}(X^TX)$ is close to zero, we can improve the condition number by shifting the spectrum $\lambda(A+\alpha I) = \lambda(A) + \alpha$, thus
\[
\kappa(X^TX+\alpha I) = \dfrac{\lambda_{\mathrm{\max}}(X^TX)+\alpha}{\lambda_{\mathrm{\min}}(X^TX)+\alpha}.
\]

\begin{figure}[H]
	\begin{tabular}{cc}
		\includegraphics[width=0.5\linewidth]{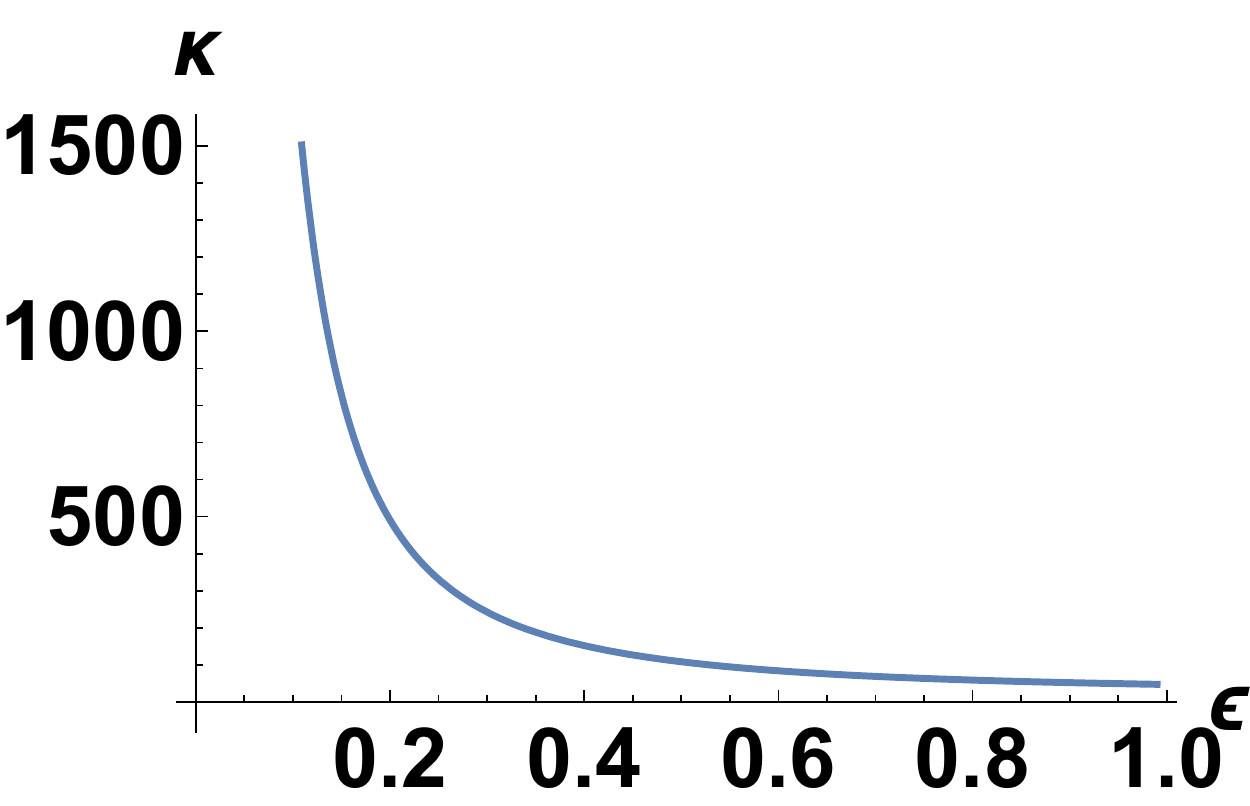} & \includegraphics[width=0.5\linewidth]{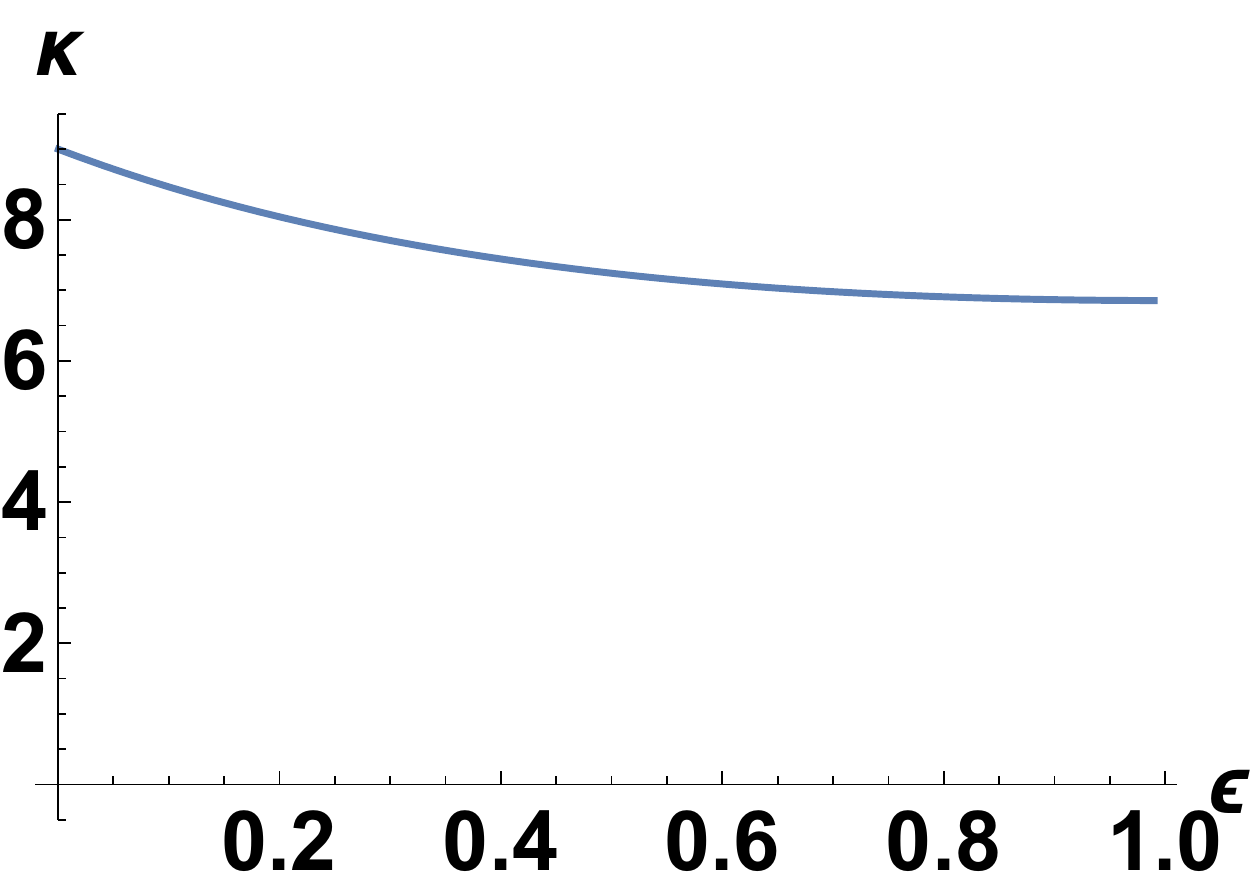}\\
		(a) $\kappa(X^TX)$ & (b) $\kappa(X^TX+I)$ 
	\end{tabular}
	\caption{Condition number of original problem (left) and the regularized one (right).}
	\label{fig:kappa}
\end{figure}

Figure~\ref{fig:kappa} compares the condition number of the original $X^TX$ matrix and the one with spectrum shifted by one $X^TX + I$.

Thus, the spectrum shift allows to address the issue of numerical instability when $ X^T X $ is ill-conditioned, which is always a case whenever $p$ is large. The solution is then given by
$$
\hat{\beta} = ( X^T X + \lambda I )^{-1} X^T y.
$$
The corresponding objective function that leads to this regularized solution is 
\begin{equation}\label{eq:ridge}
\minf_\beta ||y- X\beta||_2^2   + \lambda||\beta||_2^2.
\end{equation}

An alternative formulation is
\begin{eqnarray}\label{eq:tikhonov-constr}
\minf_\beta ||y- X\beta||_2^2   + \lambda||\beta||_2^2 \qquad\mbox{subject to}\quad ||\beta||_2^2 \le s.
\end{eqnarray}

We can think of the constrain is of a budget on the size of $\beta$. In statistics the problem of solving  (\ref{eq:ridge}) is called ridge regression. 

\subsection{Spike-and-Slab Prior}
Under spike-and-slab, prior for each $\beta_i$ is defined as a mixture of a point mass at zero, and a Gaussian distribution centered at zero
\begin{equation}
\label{eqn:ss}
\beta_i | \theta, \sigma^2\sim (1-\theta)\delta_0 + \theta N\left(0, \sigma^2\right) \ .
\end{equation}
Here $\theta\in \left(0, 1\right)$ determines the overall sparsity in $\beta$ and $\sigma^2$ accommodates non-zero signals.  This family is termed as the Bernoulli-Gaussian mixture model in the signal processing community.

A useful re-parametrization, the parameters $\beta$ is given by two independent random variable vectors $\gamma = \left(\gamma_1, \ldots, \gamma_p\right)$ and $\alpha = \left(\alpha_1, \ldots, \alpha_p\right)$ such that $\beta_i  =  \gamma_i\alpha_i$, with probabilistic structure
\begin{equation}
\label{eq:bg}
\begin{array}{rcl}
\gamma_i\mid\theta & \sim & \text{Bernoulli}(\theta) \ ;
\\
\alpha_i \mid \sigma^2&\sim & N\left(0, \sigma^2\right) \ .
\\
\end{array}
\end{equation}
Since $\gamma_i$ and $\alpha_i$ are independent, the joint prior density becomes
$$
p\left(\gamma_i, \alpha_i \mid \theta, \sigma^2\right) =
\theta^{\gamma_i}\left(1-\theta\right)^{1-\gamma_i}\frac{1}{\sqrt{2\pi}\sigma_\beta}\exp\left\{-\frac{\alpha_i^2}{2\sigma^2}\right\}
\ , \ \ \ \text{for } 1\leq i\leq p \ .
$$
The indicator $\gamma_i\in \{0, 1\}$ can be viewed as a dummy variable to indicate whether $\beta_i$ is included in the model. Under this re-parameterization, the posterior is given by

$$
\begin{array}{rcl}
-\log p\left(\gamma, \alpha \mid \theta, \sigma^2, \sigma_e^2, y\right) & \propto &
-\log p\left(\gamma, \alpha \mid \theta, \sigma^2\right)
p\left(y \mid \gamma, \alpha, \theta, \sigma_e^2\right)\\
& \propto &
\frac1{2\sigma_e^2}\left\|y - X_\gamma \alpha_\gamma\right\|_2^2
+\frac1{2\sigma^2}\left\|\alpha\right\|_2^2
+\log\left(\frac{1-\theta}{\theta}\right)
\sum_{i=1}^{p}\gamma_i.
\end{array}
$$
By construction, the $\gamma$ $\in\left\{0, 1\right\}^p$ will directly perform variable selection. Note, that the problem of minimizing the negative log-posterior is a mixed integer program with each  $\gamma_1$ being constraint to take values 0 or 1. This optimization problem is NP-hard, e.g. we cannot solve it efficiently for any meaningful value of $p$. Efficient algorithms for MAP estimation for  high dimensional linear models  were proposed in~\cite{moran2018variance,rovckova2018spike}. A sampling algorithm was proposed in ~\cite{atchade2018regularization} For a recent review of sampling algorithms for spike--and-slab, see~\cite{rockova2017dynamic}.

\section{Horseshoe}
In a global-local class of priors, $\tau$ does not depend on index $i$, therefore we have
\[
\beta_i \mid \lambda_i,\tau \sim N(0,\tau^2\lambda_i^2).
\]
Global hyper-parameter $\tau$ shrinks all parameters towards zero, while the prior for the local parameter $\lambda_i$ has a tail that decays slower than an exponential rate, and thus allows $\beta_i$ not to be shrunk. A particular representative of global-local shrinkage prior is horseshoe, which assumes half-Cauchy distribution over $\lambda_i$ and $\tau$
\[
\lambda_i \sim C^+(0,1),~~~\tau\sim C^+(0,1).
\]
Being constant at the origin, the half-Cauchy prior has nice risk properties near the origin~\citep{polson2009alternative}. \cite{polson2010shrink} warn against using empirical-Bayes or cross-validation approaches to estimate $\tau$, due to the fact that MLE estimate of $\tau$ is always in danger of collapsing to the degenerate $\hat \tau  = 0$~\citep{tiao1965bayesian}.

A feature of the horseshoe prior is that it possesses both tail-robustness and sparse-robustness properties~\citep{bhadra2017horseshoe+}; meaning that  an infinite spike at the origin and very heavy tail  that still ensures integrability. The horseshoe prior can also be specified as
\[
\beta_i\mid\lambda_i,\tau \sim N(0,\lambda_i^2),~~~\lambda_i\mid \tau \sim C^+(0,\tau),~~~\tau\sim C^+(0,1)
\]

%https://arxiv.org/pdf/1702.07400.pdf
%https://betanalpha.github.io/assets/case_studies/bayes_sparse_regression.html
%https://www2.stat.duke.edu/courses/Fall17/sta721/lectures/Lasso/lasso.pdf

The log-prior of the horseshoe cannot be calculated analytically, but a tight lower bound  ~\citep{carvalho2010horseshoe} can be used instead
\begin{equation}
\phi_{HS} ( \beta_i | \tau ) = - \log p_{HS} ( \beta_i | \tau ) \ge - \log \log \left ( 1 + \frac{2 \tau^2}{\beta_i^2} \right ) .
\end{equation}
The motivation for the horseshoe penalty arises from the analysis of the prior mass and influence on the posterior in {\bf both} the tail and behaviour at the origin. The latter provides the key determinate of the sparsity properties of the estimator.

\begin{figure}[H]
	\begin{tabular}{cc}
		\includegraphics[width=0.5\linewidth]{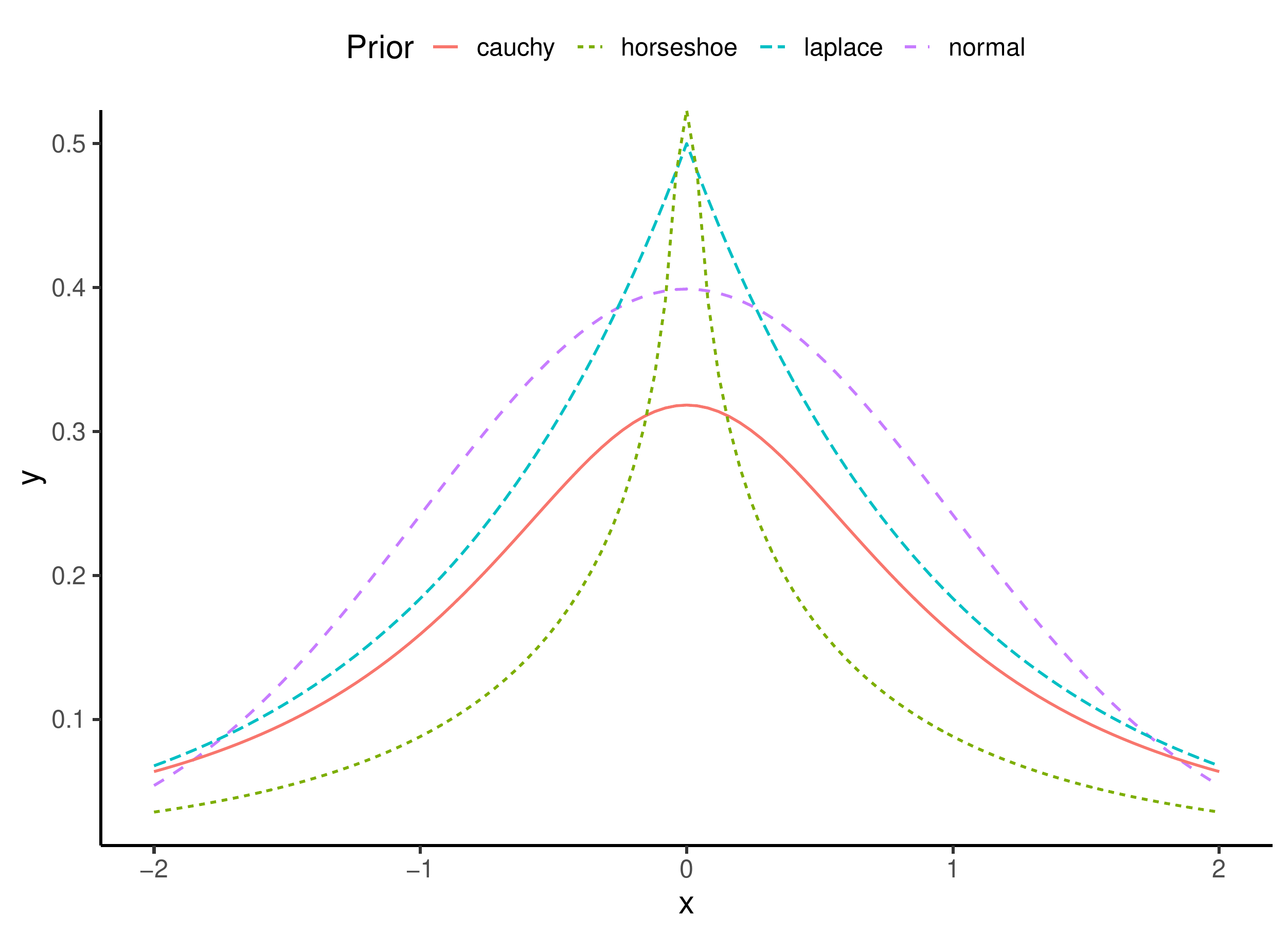} & \includegraphics[width=0.5\linewidth]{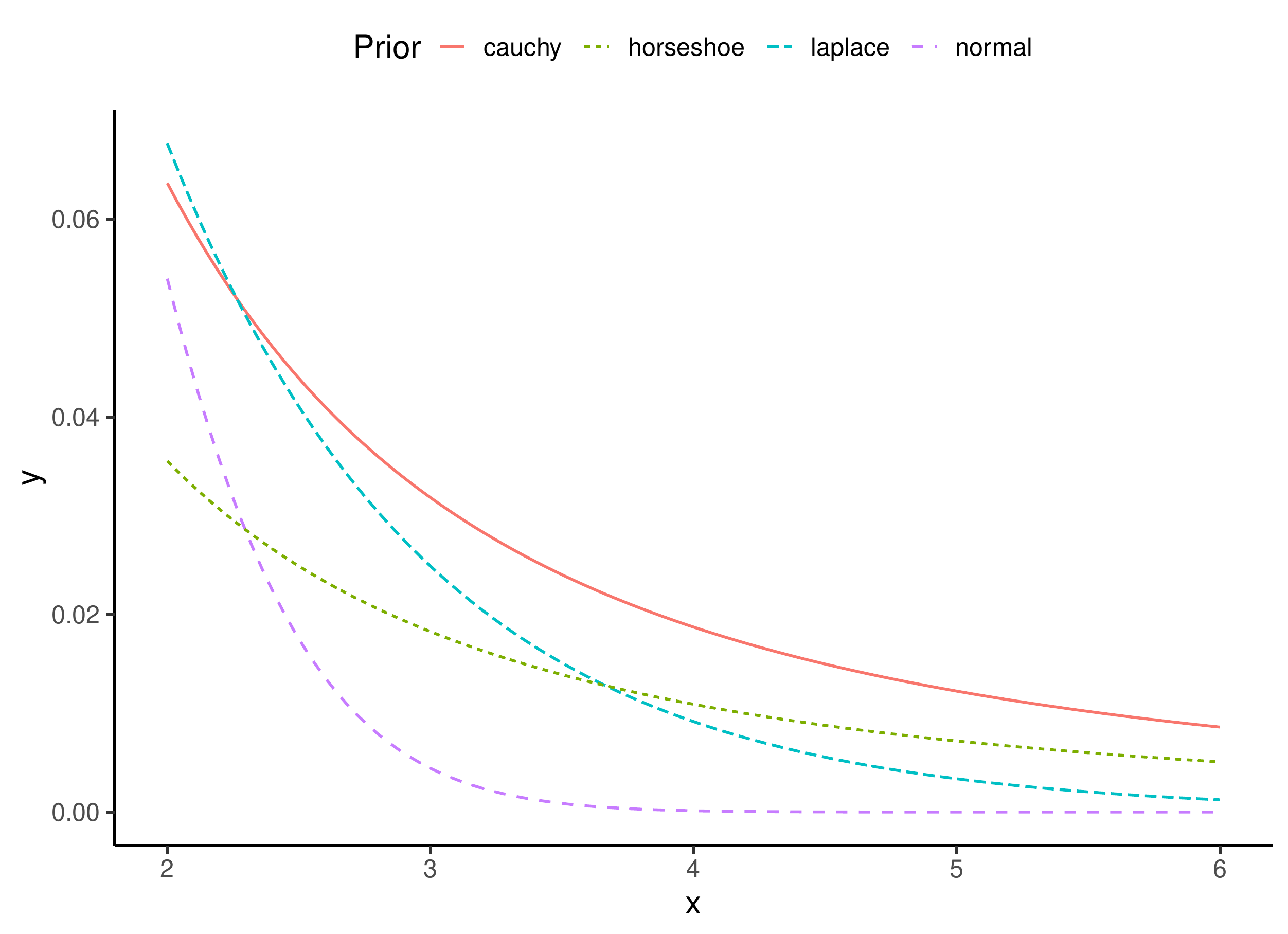}
	\end{tabular}
	\caption{Comparison of Laplace (LASSO), Normal (Ridge), Cauchy and Horseshoe priors}
	\label{fig:}
\end{figure}

When Metropolis-Hasting MCMC is applied to horseshoe regression, it suffers from sampling issues. The funnel shape geometry of the horseshoe prior is makes it challenging for MCMC to efficiently explore the parameter space. \cite{piironen2017sparsity} proposed to replace Cauchy prior with half-t proipr with small degrees of freedom and showed improved convergence behavior for NUTS sampler~\cite{hoffman2014no}. \cite{makalic2016simple} proposed using a scale mixture representation of half-Cauchy which leads to conjugate hierarchy and allows a Gibbs sample to be used.  \cite{johndrow2017scalable} proposed two MCMC algorithms to calculate posteriors for horseshoe  priors. The first algorithm addresses computational cost problem in high dimensions by approximating matrix-matrix multiplication operations. For further details on computational issues and packages for horseshoe sampling, see~\cite{bhadra2017lasso}. An issue of high dimensionality was also addressed by~\cite{bhattacharya2016fast}.

One approach is to replace the thick-tailed half-Cauchy prior over $\lambda_j$ with half-t priors using small degrees of freedom. This leads to the the sparsity-sampling efficiency trade-off problem. Larger degrees of freedom for a half-t distribution will lead to more efficient sampling algorithms, but will be less sparsity inducing. For cases with large degrees of freedom, tails of half-t are slimmer and we are required to choose large $\tau$ to accommodate large signals. However, priors with a large $\tau$ are not able to shrink coefficients towards zero as much.

\section{Empirical Results}
We use the half-Cauchy priors and a slice sampler for Bayesian linear regression models proposed in~\cite{hahn2018efficient} and implemented in the \verb|bayesml| package. The sampler does not rely on latent variables and it is automated, so that it can work with any prior that can be evaluated up to a normalizing constant. The \verb|bayeslm| package uses an elliptical slice sampler and can efficiently handle high dimensional problems. It besides horseshoe priors it also supports and spike-and-slab priors.

We then apply the slice sample to a synthetic data set. This data is generated by setting $\beta = (2,2.5,3,0,0,0,0,0,0,0)$, generating matrix $X\in R^{100\times 10}$ of 100 samples, with each uniformly distributed in $[-1,1]$. We also set $y = X\beta + e$ with $e_i\sim N(0,\kappa^2 ||\beta||_2^2)$ with $\kappa = 1$.

\begin{figure}[H]
	\centering
	\includegraphics[width=1\linewidth]{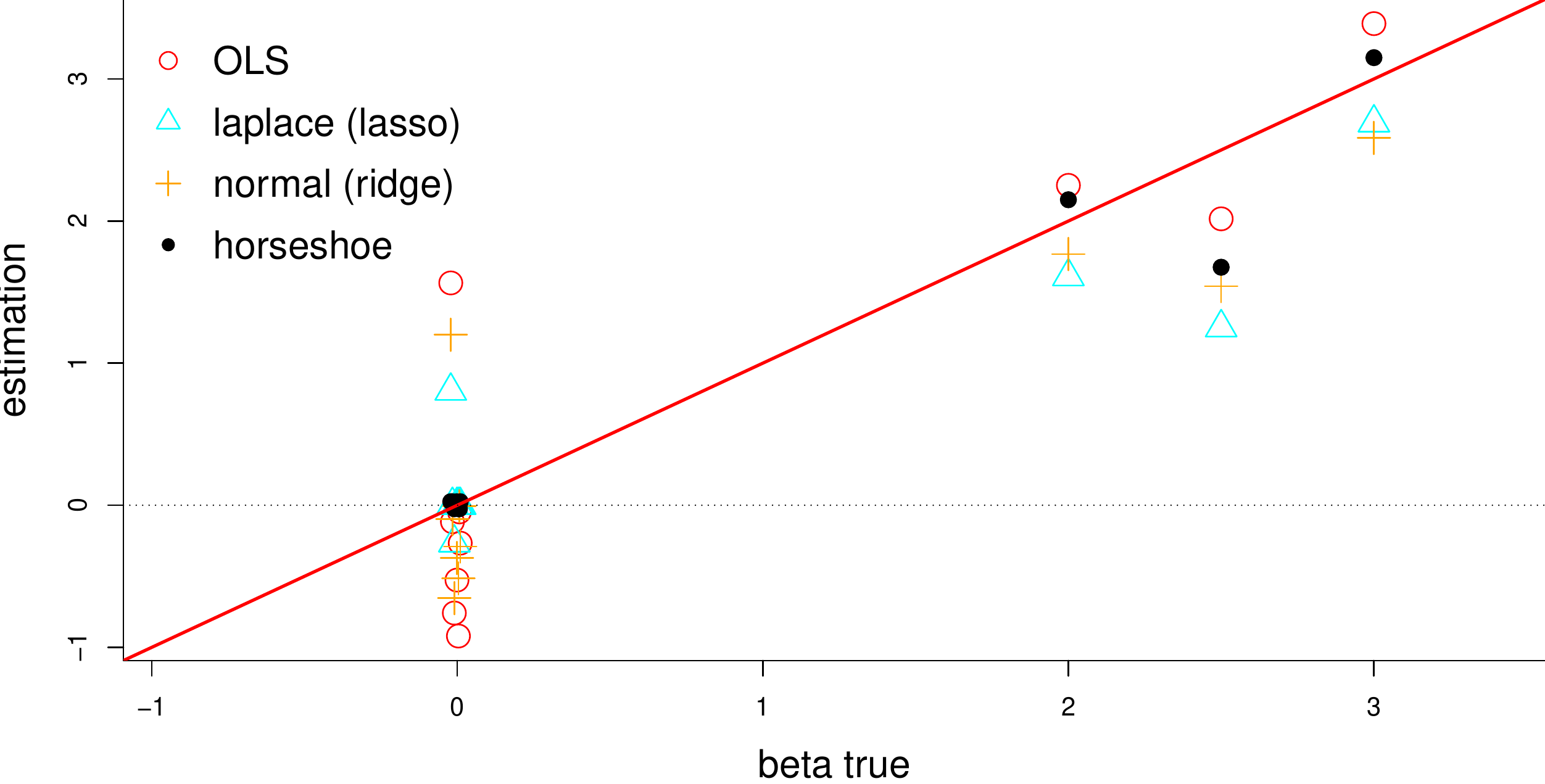}
	\caption{Posterior mode for each of each of the 10 betas estimated using Laplace, normal and horseshoe Bayesian models as well as OLS estimates. }
	\label{fig:posteriro}
\end{figure}

Figure~\ref{fig:posteriro} shows the MAP estimates using different prior assumptions as well as ordinary least squares (OLS) estimated coefficients. We can see that horseshoe was the only approach to correctly identify all zero-valued coefficients. Non-zero coefficients were recovered with a similar level of accuracy by all four methods, but we can see the shrinkage effect of the lasso estimator. 

\section{Conclusion}
There are several major advantages to using the Bayesian approach compared to the classical regularization method:
\begin{itemize}
	\item It allows for a more flexible set of models that closely match the data generating process, and assumptions appear explicitly in the model.
	\item Bayesian sampling algorithms are flexible enough and existing libraries can easily handle a wide range of model formulations without the need to design custom algorithms and implementations
	\item  Bayesian estimates are optimal  on the bias-variance scale. The parameters of the prior distribution (penalty function parameters) can be estimated using the training data set $(X,y)$~\citep{Kitagawa85} rather using brute-force search.
	\item Bayesian estimation procedures result in distributions over parameters and enable improved analysis of uncertainty in estimates and predictions. 
	\item Ability to incorporate prior information  based on expert opinion or previously observed data.
	
\end{itemize}
%First,. Second, a more flexible set of priors (penalty functions) can be applied. For example, in Bayesian approach we are not restricted to constant regularization parameter and can choose individual one for each of the model's parameters. Horseshoe is one of the most prominent examples of a Bayesian prior that has several advantages when compared to the Lasso approach. 

\bibliography{ref}

\begin{thebibliography}{}

\bibitem[\protect\astroncite{Alliney}{1992}]{alliney1992digital}
Alliney, S.\leavevmode\nopagebreak\newline 1992.
\newblock Digital filters as absolute norm regularizers.
\newblock {\em IEEE Transactions on Signal Processing}, 40(6):1548--1562.

\bibitem[\protect\astroncite{Alliney and Ruzinsky}{1994}]{alliney1994algorithm}
Alliney, S. and S.~Ruzinsky\leavevmode\nopagebreak\newline 1994.
\newblock An algorithm for the minimization of mixed l/sub 1/and l/sub 2/norms
  with application to bayesian estimation.
\newblock {\em IEEE transactions on signal processing}, 42(3):618--627.

\bibitem[\protect\astroncite{Andrews and Mallows}{1974}]{andrews1974scale}
Andrews, D.~F. and C.~L. Mallows\leavevmode\nopagebreak\newline 1974.
\newblock Scale mixtures of normal distributions.
\newblock {\em Journal of the Royal Statistical Society. Series B
  (Methodological)}, Pp.~ 99--102.

\bibitem[\protect\astroncite{Aster et~al.}{2018}]{aster2018parameter}
Aster, R.~C., B.~Borchers, and C.~H. Thurber\leavevmode\nopagebreak\newline
  2018.
\newblock {\em Parameter estimation and inverse problems}.
\newblock Elsevier.

\bibitem[\protect\astroncite{Atchade and
  Bhattacharyya}{2018}]{atchade2018regularization}
Atchade, Y. and A.~Bhattacharyya\leavevmode\nopagebreak\newline 2018.
\newblock Regularization and computation with high-dimensional spike-and-slab
  posterior distributions.
\newblock {\em arXiv preprint arXiv:1803.10282}.

\bibitem[\protect\astroncite{Bakushinskii}{1967}]{bakushinskii1967general}
Bakushinskii, A.~B.\leavevmode\nopagebreak\newline 1967.
\newblock A general method of constructing regularizing algorithms for a linear
  incorrect equation in hilbert space.
\newblock {\em Zhurnal Vychislitel'noi Matematiki i Matematicheskoi Fiziki},
  7(3):672--677.
\newblock [ English translation: \textit{U.S.S.R. Comput. Math. Math. Phys.},
  7(3) (1967), pp. 279–-287].

\bibitem[\protect\astroncite{Bhadra et~al.}{2017a}]{bhadra2017horseshoe+}
Bhadra, A., J.~Datta, N.~G. Polson, B.~Willard,
  et~al.\leavevmode\nopagebreak\newline 2017a.
\newblock The horseshoe+ estimator of ultra-sparse signals.
\newblock {\em Bayesian Analysis}, 12(4):1105--1131.

\bibitem[\protect\astroncite{Bhadra et~al.}{2017b}]{bhadra2017lasso}
Bhadra, A., J.~Datta, N.~G. Polson, and B.~T.
  Willard\leavevmode\nopagebreak\newline 2017b.
\newblock Lasso meets horseshoe.
\newblock {\em arXiv preprint arXiv:1706.10179}.

\bibitem[\protect\astroncite{Bhattacharya et~al.}{2016}]{bhattacharya2016fast}
Bhattacharya, A., A.~Chakraborty, and B.~K.
  Mallick\leavevmode\nopagebreak\newline 2016.
\newblock Fast sampling with gaussian scale mixture priors in high-dimensional
  regression.
\newblock {\em Biometrika}, P.~ asw042.

\bibitem[\protect\astroncite{Cand{\`e}s and
  Wakin}{2008}]{candes2008introduction}
Cand{\`e}s, E.~J. and M.~B. Wakin\leavevmode\nopagebreak\newline 2008.
\newblock An introduction to compressive sampling [a sensing/sampling paradigm
  that goes against the common knowledge in data acquisition].
\newblock {\em IEEE signal processing magazine}, 25(2):21--30.

\bibitem[\protect\astroncite{Carlin and Polson}{1991}]{carlin1991inference}
Carlin, B.~P. and N.~G. Polson\leavevmode\nopagebreak\newline 1991.
\newblock Inference for nonconjugate bayesian models using the gibbs sampler.
\newblock {\em Canadian Journal of statistics}, 19(4):399--405.

\bibitem[\protect\astroncite{Carlin et~al.}{1992}]{carlin1992monte}
Carlin, B.~P., N.~G. Polson, and D.~S. Stoffer\leavevmode\nopagebreak\newline
  1992.
\newblock A monte carlo approach to nonnormal and nonlinear state-space
  modeling.
\newblock {\em Journal of the American Statistical Association},
  87(418):493--500.

\bibitem[\protect\astroncite{Carvalho et~al.}{2010}]{carvalho2010horseshoe}
Carvalho, C.~M., N.~G. Polson, and J.~G. Scott\leavevmode\nopagebreak\newline
  2010.
\newblock The horseshoe estimator for sparse signals.
\newblock {\em Biometrika}, 97(2):465--480.

\bibitem[\protect\astroncite{Claerbout and Muir}{1973}]{claerbout1973robust}
Claerbout, J.~F. and F.~Muir\leavevmode\nopagebreak\newline 1973.
\newblock Robust modeling with erratic data.
\newblock {\em Geophysics}, 38(5):826--844.

\bibitem[\protect\astroncite{Donoho}{1992}]{donoho1992superresolution}
Donoho, D.~L.\leavevmode\nopagebreak\newline 1992.
\newblock Superresolution via sparsity constraints.
\newblock {\em SIAM journal on mathematical analysis}, 23(5):1309--1331.

\bibitem[\protect\astroncite{Donoho and Johnstone}{1995}]{donoho1995adapting}
Donoho, D.~L. and I.~M. Johnstone\leavevmode\nopagebreak\newline 1995.
\newblock Adapting to unknown smoothness via wavelet shrinkage.
\newblock {\em Journal of the american statistical association},
  90(432):1200--1224.

\bibitem[\protect\astroncite{George and McCulloch}{1993}]{george1993variable}
George, E.~I. and R.~E. McCulloch\leavevmode\nopagebreak\newline 1993.
\newblock Variable selection via gibbs sampling.
\newblock {\em Journal of the American Statistical Association},
  88(423):881--889.

\bibitem[\protect\astroncite{Hahn et~al.}{2018}]{hahn2018efficient}
Hahn, P.~R., J.~He, and H.~F. Lopes\leavevmode\nopagebreak\newline 2018.
\newblock Efficient sampling for gaussian linear regression with arbitrary
  priors.
\newblock {\em Journal of Computational and Graphical Statistics},
  (just-accepted).

\bibitem[\protect\astroncite{Hoerl and Kennard}{1970}]{hoerl1970ridge}
Hoerl, A.~E. and R.~W. Kennard\leavevmode\nopagebreak\newline 1970.
\newblock Ridge regression: Biased estimation for nonorthogonal problems.
\newblock {\em Technometrics}, 12(1):55--67.

\bibitem[\protect\astroncite{Hoffman and Gelman}{2014}]{hoffman2014no}
Hoffman, M.~D. and A.~Gelman\leavevmode\nopagebreak\newline 2014.
\newblock The no-u-turn sampler: adaptively setting path lengths in hamiltonian
  monte carlo.
\newblock {\em Journal of Machine Learning Research}, 15(1):1593--1623.

\bibitem[\protect\astroncite{Ivanov}{1962}]{ivanov1962linear}
Ivanov, V.~K.\leavevmode\nopagebreak\newline 1962.
\newblock On linear problems which are not well-posed.
\newblock In {\em Doklady Akademii Nauk}, volume 145, Pp.~ 270--272. Russian
  Academy of Sciences.

\bibitem[\protect\astroncite{Ivanov et~al.}{2013}]{ivanov2013theory}
Ivanov, V.~K., V.~V. Vasin, and V.~P. Tanana\leavevmode\nopagebreak\newline
  2013.
\newblock {\em Theory of linear ill-posed problems and its applications},
  volume~36.
\newblock Walter de Gruyter.

\bibitem[\protect\astroncite{Johndrow et~al.}{2017}]{johndrow2017scalable}
Johndrow, J.~E., P.~Orenstein, and
  A.~Bhattacharya\leavevmode\nopagebreak\newline 2017.
\newblock Scalable mcmc for bayes shrinkage priors.
\newblock {\em arXiv preprint arXiv:1705.00841}.

\bibitem[\protect\astroncite{Kitagawa and Gersch}{1985}]{Kitagawa85}
Kitagawa, G. and W.~Gersch\leavevmode\nopagebreak\newline 1985.
\newblock A smoothness priors time-varying ar coefficient modeling of
  nonstationary covariance time series.
\newblock {\em IEEE Transactions on Automatic Control}, 30(1):48--56.

\bibitem[\protect\astroncite{Makalic and Schmidt}{2016}]{makalic2016simple}
Makalic, E. and D.~F. Schmidt\leavevmode\nopagebreak\newline 2016.
\newblock A simple sampler for the horseshoe estimator.
\newblock {\em IEEE Signal Processing Letters}, 23(1):179--182.

\bibitem[\protect\astroncite{Miller}{2002}]{miller2002subset}
Miller, A.\leavevmode\nopagebreak\newline 2002.
\newblock {\em Subset selection in regression}.
\newblock Chapman and Hall/CRC.

\bibitem[\protect\astroncite{Moran et~al.}{2018}]{moran2018variance}
Moran, G.~E., V.~Rockova, and E.~I. George\leavevmode\nopagebreak\newline 2018.
\newblock Variance prior forms for high-dimensional bayesian variable
  selection.
\newblock {\em arXiv preprint arXiv:1801.03019}.

\bibitem[\protect\astroncite{Park and Casella}{2008}]{park2008bayesian}
Park, T. and G.~Casella\leavevmode\nopagebreak\newline 2008.
\newblock The bayesian lasso.
\newblock {\em Journal of the American Statistical Association},
  103(482):681--686.

\bibitem[\protect\astroncite{Piironen et~al.}{2017}]{piironen2017sparsity}
Piironen, J., A.~Vehtari, et~al.\leavevmode\nopagebreak\newline 2017.
\newblock Sparsity information and regularization in the horseshoe and other
  shrinkage priors.
\newblock {\em Electronic Journal of Statistics}, 11(2):5018--5051.

\bibitem[\protect\astroncite{Polson and Scott}{2009}]{polson2009alternative}
Polson, N.~G. and J.~G. Scott\leavevmode\nopagebreak\newline 2009.
\newblock Alternative global--local shrinkage rules using hypergeometric--beta
  mixtures.
\newblock {\em Technical report 14}.

\bibitem[\protect\astroncite{Polson and Scott}{2010}]{polson2010shrink}
Polson, N.~G. and J.~G. Scott\leavevmode\nopagebreak\newline 2010.
\newblock Shrink globally, act locally: Sparse bayesian regularization and
  prediction.
\newblock {\em Bayesian statistics}, 9:501--538.

\bibitem[\protect\astroncite{Polson and Sun}{2017}]{polson2017bayesian}
Polson, N.~G. and L.~Sun\leavevmode\nopagebreak\newline 2017.
\newblock Bayesian l 0-regularized least squares.
\newblock {\em Applied Stochastic Models in Business and Industry}.

\bibitem[\protect\astroncite{Ro{\v{c}}kov{\'a} and
  George}{2018}]{rovckova2018spike}
Ro{\v{c}}kov{\'a}, V. and E.~I. George\leavevmode\nopagebreak\newline 2018.
\newblock The spike-and-slab lasso.
\newblock {\em Journal of the American Statistical Association},
  113(521):431--444.

\bibitem[\protect\astroncite{Rockova and McAlinn}{2017}]{rockova2017dynamic}
Rockova, V. and K.~McAlinn\leavevmode\nopagebreak\newline 2017.
\newblock Dynamic variable selection with spike-and-slab process priors.
\newblock {\em arXiv preprint arXiv:1708.00085}.

\bibitem[\protect\astroncite{Stein}{1964}]{stein1964inadmissibility}
Stein, C.\leavevmode\nopagebreak\newline 1964.
\newblock Inadmissibility of the usual estimator for the variance of a normal
  distribution with unknown mean.
\newblock {\em Annals of the Institute of Statistical Mathematics},
  16(1):155--160.

\bibitem[\protect\astroncite{Taylor et~al.}{1979}]{taylor1979deconvolution}
Taylor, H.~L., S.~C. Banks, and J.~F. McCoy\leavevmode\nopagebreak\newline
  1979.
\newblock Deconvolution with the $\ell_1$ norm.
\newblock {\em Geophysics}, 44(1):39--52.

\bibitem[\protect\astroncite{Tiao and Tan}{1965}]{tiao1965bayesian}
Tiao, G.~C. and W.~Tan\leavevmode\nopagebreak\newline 1965.
\newblock Bayesian analysis of random-effect models in the analysis of
  variance. i. posterior distribution of variance-components.
\newblock {\em Biometrika}, 52(1/2):37--53.

\bibitem[\protect\astroncite{Tibshirani}{1996}]{tibshirani1996regression}
Tibshirani, R.\leavevmode\nopagebreak\newline 1996.
\newblock Regression shrinkage and selection via the lasso.
\newblock {\em Journal of the Royal Statistical Society. Series B
  (Methodological)}, Pp.~ 267--288.

\bibitem[\protect\astroncite{Tihonov}{1963}]{tihonov1963solution}
Tihonov, A.~N.\leavevmode\nopagebreak\newline 1963.
\newblock Solution of incorrectly formulated problems and the regularization
  method.
\newblock {\em Soviet Math.}, 4:1035--1038.

\bibitem[\protect\astroncite{Tikhonov and Arsenin}{1977}]{tikhonov1977methods}
Tikhonov, A. and V.~Y. Arsenin\leavevmode\nopagebreak\newline 1977.
\newblock {\em Methods for solving ill-posed problems}.
\newblock John Wiley and Sons, Inc.

\bibitem[\protect\astroncite{Tikhonov}{1943}]{tikhonov1943stability}
Tikhonov, A.~N.\leavevmode\nopagebreak\newline 1943.
\newblock {O}n the stability of inverse problems.
\newblock In {\em Dokl. Akad. Nauk SSSR}, volume~39, Pp.~ 195--198.

\bibitem[\protect\astroncite{Williams}{1995}]{williams1995bayesian}
Williams, P.~M.\leavevmode\nopagebreak\newline 1995.
\newblock Bayesian regularization and pruning using a laplace prior.
\newblock {\em Neural computation}, 7(1):117--143.

\end{thebibliography}
\end{document}